# Massive enhancement of electron-phonon coupling in doped graphene by an electronic singularity

Jessica L. McChesney,[1-3] Aaron Bostwick,[1] Taisuke Ohta,[1,3] Konstantin V. Emtsev,[4] Thomas Seyller,[4] Karsten Horn,[3] and Eli Rotenberg[1]

[1]Advanced Light Source, E. O. Lawrence Berkeley National Laboratory, Berkeley, CA 94720 USA

[2]Department of Physics, Montana State University, Bozeman MT, 59717

[3]Department of Molecular Physics, Fritz-Haber-Institut der Max-Planck-Gesellschaft, Faradayweg 4-6, 14195 Berlin, Germany

[4]Lehrstuhl für Technische Physik, Universität Erlangen-Nürnberg, Erwin-Rommel-Straße 1, D-91058 Erlangen, Germany

## I. Abstract

The nature of the coupling leading to superconductivity in layered materials such as high-$T_c$ superconductors and graphite intercalation compounds (GICs) is still unresolved. In both systems, interactions of electrons with either phonons or other electrons or both have been proposed to explain superconductivity.[1-3] In the high-$T_c$ cuprates, the presence of a Van Hove singularity (VHS) in the density of states near the Fermi level was long ago proposed to enhance the many-body couplings and therefore may play a role in superconductivity.[4] Such a singularity can cause an anisotropic variation in the coupling strength, which may partially explain the so-called nodal-antinodal dichotomy[5] in the cuprates. Here we show that the topology of the graphene band structure at dopings comparable to the GICs is quite similar to that of the cuprates and that the quasiparticle dynamics in graphene have a similar dichotomy. Namely, the electron-phonon coupling is highly anisotropic, diverging near a saddle point in the graphene electronic band structure. These results support the important role of the VHS in layered materials and the possible optimization of $T_c$ by tuning the VHS with respect to the Fermi level.

## II. Introduction

A complete understanding of the nature of superconductivity in layered compounds such as high-$T_c$ superconductors (HTSs) and graphite intercalation compounds (GICs) is still elusive despite an intense effort.

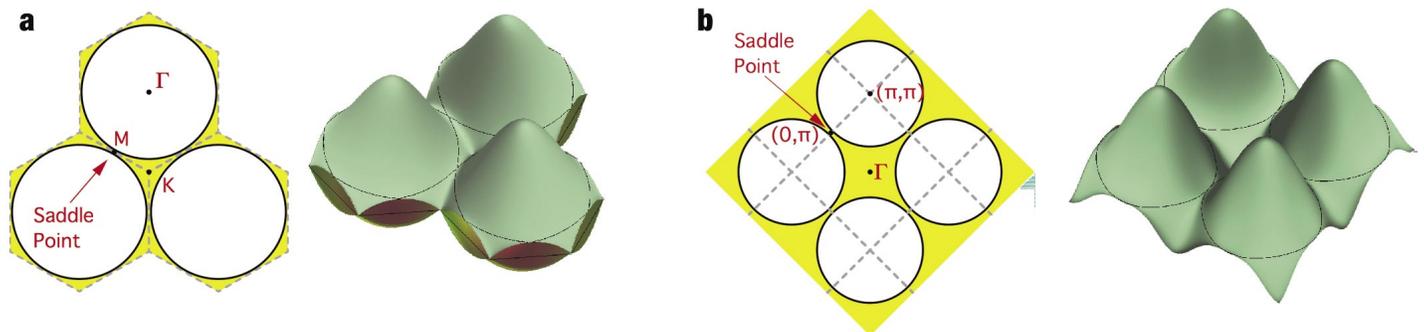

**Figure 1**. **Saddle-point band structure for 2D metals.** Model Fermi surfaces and Brillouin Zones (upper) and third-neighbor hopping model band structures (lower) for **a** doped graphene π* bands[24] and **b** hole-doped High-$T_c$ compounds[35]. Both show saddle points (van Hove Singularities) at the intersections of the circular contours near $E_F$.

A central issue is the relative strength of electron-electron (*e-e*) and electron-phonon (*e-ph*) interactions.[1-3] There is a notable similarity of the GIC and normal-state HTS Fermi surfaces which both consist of nearly circular hole pockets whose underlying bands have a saddle-point topology at their intersections (Fig. 1).[6, 7] Such a band topology is important because the electronic density of states at such a saddle point diverges logarithmically, and it has been predicted that such singularities (also called van Hove singularities, or VHSs) can strongly influence the many-body interactions if they are situated at or near the Fermi energy $E_F$.[4, 8] These VHSs occur along the KM direction for graphene π* bands and along the (0,π) direction for HTSs. Correspondingly, in the HTSs, the character of the quasiparticles (QPs) along the (0,π) and (π,π) ("anti-nodal" and "nodal") directions have a famous "dichotomy"[5]: Along the nodal direction the QPs are well-defined, while in the anti-nodal directions near the saddle points, the QPs are either broad, or else completely incoherent depending on doping. Here we demonstrate a similar dichotomy for doped graphene.

The importance of the VHS for superconductivity is that it is associated with divergences in the spin and charge susceptibilities, so that the *e-e* and *e-ph* interactions can be greatly enhanced. Whether or not this enhancement drives the coupling in high temperature superconductivity (this situation is called the "VHS scenario"[4]), one can argue that the effect of the VHS must be part of any relevant theory of high-$T_c$ superconductivity. Despite the undisputed presence of the VHS in the cuprates,[6, 7] the VHS scenario has gone in and out of favour over the years.

In this context we present a systematic angle-resolved photoemission spectroscopy (ARPES) study of an individually doped graphene plane, an elegant model system to study the VHS scenario. We find a surprisingly strong and anisotropic enhancement of *e-ph* coupling strength upon tuning the Fermi level through a VHS. Since graphene can be prepared with few defects and is not known to have complications such as density waves, stripes, and magnetic excitations, it is likely that the VHS causes the enhanced coupling observed in graphene. Our findings also suggest that the variation in $T_c$ among the GICs (from <1 to 11.5K depending on the intercalant[9-11]) may be influenced by the vicinity of a VHS to the Fermi level.

### III. Experimental

The ARPES experiments were conducted at the Electronic Structure Factory endstation at beamline 7.01 at the Advanced Light Source using a Scienta R4000 electron analyser. Band maps were taken at T < 30K with $h\nu$ = 94 eV photons with an overall resolution of ~ 25 meV. The base pressure was < 2.5 × 10$^{-11}$ Torr, ensuring long sample lifetimes.

While isolation of single layer graphene flakes has been demonstrated by exfoliation,[12, 13] in our experiments, clean macroscopic graphene surfaces (Fig. 2a) were prepared by annealing SiC and characterized as described previously.[14-19] Fig. 2 shows the evolution of the graphene π* Fermi surfaces and the underlying bands as a function of doping by various combinations of K and Ca, starting from clean graphene. The clean bands, with a linear dispersion around the Dirac energy $E_D$ (Fig. 2a), are shifted by charge transfer from the substrate such that $E_F - E_D$=0.45 eV and have an approximately circular Fermi surface centered at the

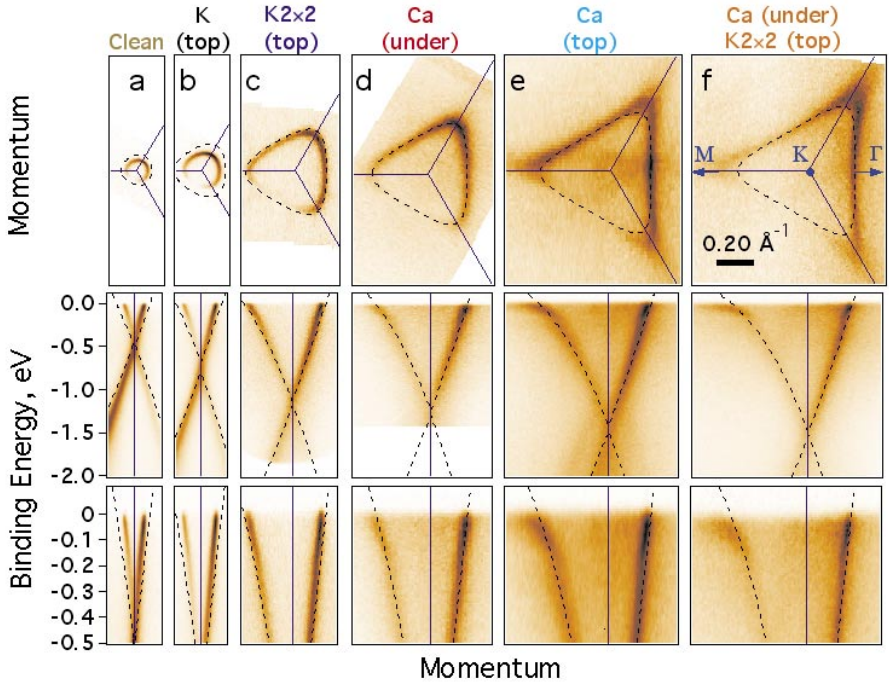

**Figure 2. Doping-dependence of graphene's electronic structure.** Doping dependence of (top) Fermi surface and (middle, bottom) bandstructure of π bands along the MKΓ directions, for various combinations of doping from (a) as-grown graphene, (b-f) graphene with Ca and/or K deposition as labeled. The DFT calculated bands (as fitted by third-NN TB bands) and Fermi surfaces[24] for graphene are shown by dashed lines.

K-point (Fig 2a, top).[16, 17] Upon doping with electrons (achieved by depositing combinations of K and Ca layers), the Fermi surface grows in size, its slight trigonal warping evolving into a distinctly concave triangular shape (Fig. 2b-f, top).

At low coverages and temperature, K adsorbs in a dispersed, disordered fashion as described previously for graphene [17] and graphite [20], while deposition of K at T~100°K results in an ordered K 2×2 reconstruction, stoichiometrically equivalent to bulk $KC_8$ (Fig. 2(b,c)). In the dispersed phase, no overlayer state could be observed but in the K 2×2 phase, a circular contour around the Γ point was observed, representing overlayer K states.

In agreement with Ca growth on Cu(001), we observed no low-coverage, dispersed Ca phase. Instead, Ca deposition at low temperature followed by annealing leads to ordered c(1×1) (Fig. 2(e)) or (√3 × √3)R30° (henceforth √3 ×) phases, for coverages less than or just greater than 1 ML, respectively. These phases have very similar Fermi surfaces consisting of triangular π* bands centered on the K-point, and a central, circular contour attributed to overlayer, Ca-derived states. (The √3 × phase is characterized by weak backfolding of the bands with √3 × symmetry.) It appears that the long-range √3 × order is stabilized by a small amount of extra Ca beyond the first layer, but otherwise the Fermi surfaces of the π* and Ca overlayer bands in the first layer are equivalent. Data in the paper are shown for the c(1×1) phase.

The Fermi surfaces we observed are therefore very similar to bulk $KC_8$ and $CaC_6$, except that the spherical interlayer Fermi surfaces centered at Γ are replaced by 2D cylindrical sheets. This is not surprising considering the layered nature of bulk $KC_8$ and $CaC_6$. From the Fermi surface areas of the π* and the overlayer states, the doping levels could be determined which confirms that K 2×2 and Ca-√3 × phases are consistent with the charge expected for 1/8 and 1/6 ML respectively.

We also found we could intercalate the Ca between the graphene and SiC. Intercalation under the graphene was possible only for repeated cycles of Ca deposition and annealing onto slightly incomplete graphene layers, which we believe consist of isolated, but nearly merged, islands of monolayer graphene. This procedure is similar to the promotion of bulk $CaC_6$ formation by the presence of grain boundaries in the graphene planes which presumably provide a path for Ca intercalation [21, 22]. The phases Ca(under) and Ca(top) could be distinguished because the former was stable up to 1100°C, while the latter was easily re-evaporated at only a few hundred degrees C. Furthermore, the Fermi surface of the underlayer phase (Fig. 2(d)) was smaller than the overlayer phase, indicating either a less dense phase, or else less charge transfer to the π* band.

### IV. Bandstructure Findings

We now consider the details of the band structures (center and bottom rows of Fig. 2). Apart from minor deviations in the band velocity $v \equiv \partial E/\partial k$ (attributable to a small change in lattice constant with doping[23]), the bands more or less follow Reich's tight binding (TB) fits to density functional bands[24] (dashed lines). Closer to $E_F$ (bottom panels), we find more profound deviations. For relatively low doping, the bands are kinked at ~200 meV below $E_F$ due to *e-ph* coupling to the graphene in-plane optical vibrations[17] similar to bilayer graphene[18] and graphite.[25, 26] This *e-ph* kink becomes strikingly apparent at higher doping in the KM direction, becoming so strong as to bend the bands well outside Reich's calculation, which we take to represent the bare, unrenormalized bands. The abruptness of the kink, both in terms of a change of band velocity as well as a broadening of the bands below $E_F$, cannot easily be explained by single-particle effects and suggests a strong mass renormalization by electron-phonon coupling.

We extract the *e-ph* coupling constant $\lambda_\pi(\mathbf{k})$ from the measured slopes of the π* band near $E_F$ as follows. The band velocity near $E_F$ is renormalized by electron-phonon coupling. For an Einstein phonon mode it is renormalized by the factor $(1+\lambda_\pi(\mathbf{k}))^{-1}$, whence $\lambda_\pi(\mathbf{k})=v_0/v_F -1$. Here $\lambda_\pi(\mathbf{k})$ is the momentum-dependent *e-ph* coupling constant between C π* electrons and the C optical phonon modes, and $v_0$ is the bare band velocity (in the absence of *e-ph* coupling), taken as the slope of the bands below the phonon kink at ~200 meV. By fitting the dispersions of the band slices taken radially outward from the K point, we extracted the band velocities $v_0$ and $v_F$ as a function of azimuthal angle, thus determining $\lambda_\pi(\mathbf{k})$. This procedure was reliable for all but the lowest doping, in which there was overlap between the *el-ph* kink and a kink at $E_D$ due to electron-plasmon[17] coupling. Therefore Fig. 3 omits coupling data for clean graphene.

The Fermi contours, derived from curvefitting the data in Fig. 2 for the various dopings, are plotted in Fig. 3a with coloring according to these experimentally determined $\lambda_\pi(\mathbf{k})$ values. These $\lambda_\pi(\mathbf{k})$ values for each doping are also shown in Fig. 3b as polar plots; as guides to the eye, we show fits to the data by an empirical function (solid lines). These plots reveal that $\lambda_\pi(\mathbf{k})$ has a dramatic anisotropy along each Fermi contour (around 5:1 ratio at the highest doping), with maxima along the KM lines of the Brillouin zone. To our knowledge, the only comparable anisotropy within a single band was predicted for the coupling of oxygen vibrations to electrons in HTSs[8], ascribed not only to a VHS but other factors as well. Fig. 3c shows the maximum, minimum, and angle-averaged coupling parameter *vs.* doping. While the minimum coupling strength grows only slowly with doping, the maximum coupling parameter diverges as the corresponding segment of the Fermi contour approaches the VHS at the M point.

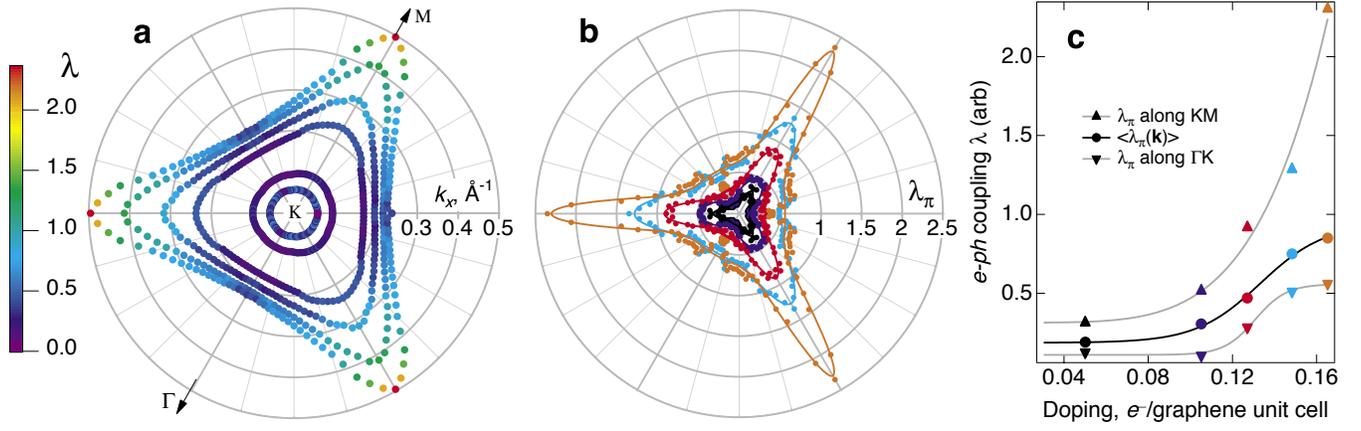

**Figure 3. Doping dependence of $\lambda_\pi(\mathbf{k})$.** **a** the symmetrized Fermi contours as a function of doping, colored according to $\lambda_\pi(\mathbf{k})$. **b** polar plot of $\lambda_\pi(\mathbf{k})$ about one of the K points (markers) fitted to an empirical function as a guide to the eye (lines) for various dopings showing a high degree of anisotropy. **c** maximal and minimal values of $\lambda_\pi(\mathbf{k})$ (along KM and ΓK directions, resp.), and momentum-averaged value $\langle\lambda_\pi(\mathbf{k})\rangle$. The doping levels in **b** and **c** are the same as the 5 highest dopings in Fig. 2 and have a common color palette.

Such a strong signature of *e-ph* coupling is quite unexpected, not only because of its strength (expected to be weak for π-bands and optical phonons in graphene[3] and CaC$_6$[3, 11, 27, 28]), but also because of its anisotropy, which cannot be explained by simple *e-ph* coupling models or density functional theory (DFT) calculations[3, 28, 29]. Our own calculations of the *el-ph* coupling matrix elements indicate only a weak anisotropy, if any.

The presence of a strong VHS along the KMK zone boundary is confirmed upon further doping. In Fig. 4a-b the Fermi surface and bands are shown for a similar doping level as in Fig. 2f, achieved through a combination of Ca atoms placed both above and below the graphene. We can see a hint of intensity at $E_F$ connecting the corners of adjacent Fermi surfaces in Fig. 4a; we attribute this to the tail of an unoccupied band just above $E_F$. This apparently flat band, when filled by doping with additional K atoms, falls below $E_F$ (Fig. 4c-d), bridging the kinked bands along the KMK line. The notable flatness of this band signifies an *extended* VHS, in similarity to the cuprates,[6, 7] which we believe causes the divergence of $\lambda_\pi(\mathbf{k})$ along the Brillouin zone boundary in graphene. While this flatness may originate in the single-particle bandstructure (namely a Ca *d* character of the band[3]), it is most likely further enhanced by the *e-ph* coupling. The presence of a "peak-dip-hump" lineshape along the entire extended VHS line (Fig. 4e) confirms that the flat band is strongly coupled to phonons along its entire length. The extended nature of the VHS is important because it converts the logarithmic divergence in the density of states to a much stronger square-root divergence, potentially further enhancing $T_c$ in the VHS scenario.[4]

## V. Discussion and Conclusions

It is tempting to relate our findings to superconductivity in bulk (3-dimensional) GICs KC$_8$ and CaC$_6$. For our K(2×2) layer (Fig. 2(c)), the atomic arrangement as well as π* bandstructure closely match those of bulk $KC_8$[30] which has an inferred phonon coupling constant[31] $\lambda \sim 0.3$. This is in close agreement with the angle-averaged $\langle \lambda_\pi(\mathbf{k}) \rangle$ determined for our K(2×2) layer. Furthermore, we observe an additional circular Fermi contour centered at Γ —associated with the overlayer potassium atoms— which has no apparent kink at any energy scale down to our resolution (~25 meV). This suggests that the dominant *e-ph* coupling in K(2×2), and by extension $KC_8$, is between C π* electrons and C in-plane vibrations, slightly enhanced due to the proximity of the VHS above $E_F$.

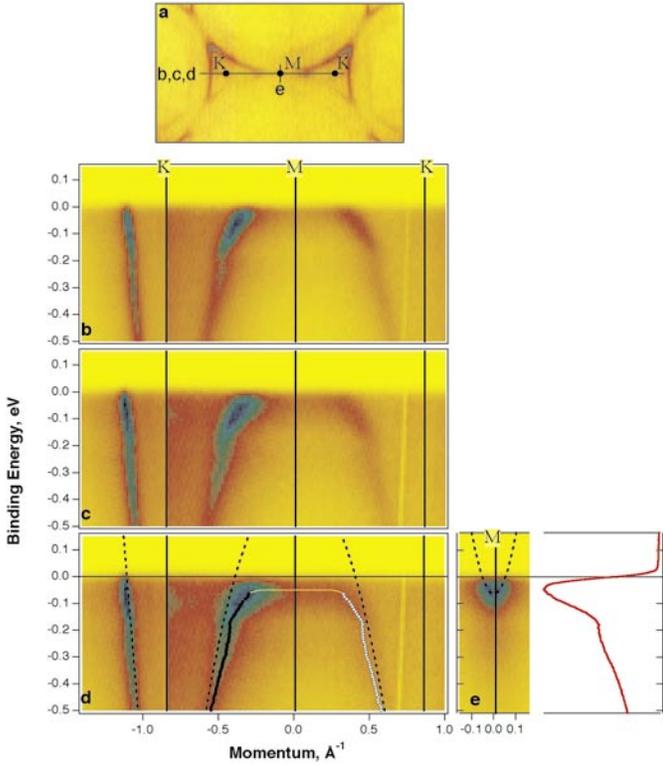

**Figure 4. Extended van Hove Singularity in Graphene. a** Fermi surface and **b-e** Energy bands along the lines indicated in **a**. **b** Bands for a doping similar to Fig. 2f. **c-d** Bands for progressively higher dopings, showing the emergence of a flat band near $E_F$. The band positions away from $E_F$ are shown by the circles (solid=as fitted to momentum distribution curves, hollow=symmetrized); the yellow line is a guide to the eye to highlight the flat band near $E_F$. **e** Angle-resolved and angle-integrated bands through the M point for the same doping as **d**. The angle-integrated spectrum shows a peak-dip-hump structure which is also faintly seen in the main panels. The dashed lines are the DFT calculated bands[24] as fitted by a third-NN TB model and shifted in energy to align with the data.

A comparison of our higher-doped bands to $CaC_6$ is less straightforward, as there are no bulk measurements of the $CaC_6$ bands and our Ca atoms may not have the same atomic arrangement as in $CaC_6$. Nevertheless at the approximate $CaC_6$ stoichiometry (Fig. 2(e)) we find for Ca on graphene an average coupling constant $\langle \lambda_\pi(\mathbf{k}) \rangle = 0.85$, in agreement with the inferred value from specific heat measurements[32] and first principles calculations[27-29] for bulk $CaC_6$. This suggests the possibility that $CaC_6$ superconductivity is also dominated by C π* electrons, whose coupling is now greatly enhanced by the presence of the VHS, in contrast to these calculations.

By extension, our findings support the idea that the VHS is also important for the HTSs where it is debated whether either electron-electron, electron-phonon or other interactions determine the total coupling responsible for superconductivity.[1] While it is possible that the VHS enhances the *el-ph* (or other bosonic) coupling directly,[4, 8] *el-el* interactions may also come into play. For example, at a saddle point, there is a mixture of electron and hole character which can lead to an attractive Coulomb interaction.[4] Such a landscape of multiple interactions was proposed to explain the unusual isotope effects on the spectral function of HTSs[33] and may also be the situation in doped graphene. Furthermore, the appearance of an extended VHS in both systems [6, 7] suggests the emergence of a one-dimensional character to the electronic states;[34] how this change in dimensionality is related to the enhancement of *el-ph* coupling should be the subject of further theoretical work.

Our analysis of the influence of many-body effects supports the VHS scenario that electron-boson coupling and hence superconductivity in GICs and other layered compounds may be greatly enhanced if, by tuning the composition, a VHS can be placed exactly at $E_F$. This should also hold for *p*-doped GICs, as there is a similar VHS below the Dirac crossing energy which could be reached in principle by S, Cl, or other *p* dopants. Finally, considering the ease with which thin graphene films can be doped in a gated geometry, it suggests the possibility of a graphene-based superconducting device whose superconductivity can be electrically controlled.

## VI. Acknowledgements

We are grateful to M. Calandra and F. Mauri for discussions, and to P. D. Johnson, R. Markiewicz, and T. P. Devereaux for comments on the manuscript. Research at the Advanced Light Source was supported by the Director, Office of Science, Office of Basic Energy Sciences, of the U.S. Department of Energy. J. M., T. O., and K. H. were supported by the Max Planck Society and the European Science Foundation under the EUROCORES SONS program.